\documentclass[12pt,preprint]{aastex}
\usepackage{geometry}
\usepackage{graphics}
\usepackage{graphicx}
\title{Single- and Two-Component GRB Spectra \\
in the Fermi GBM-LAT Energy Range}
\author{P. Veres and P. \Mesz}
\affil{Dept. of Astronomy \& Astrophysics, Dept. of Physics and Center for
Particle Astrophysics, 525 Davey Lab.,
Pennsylvania State University, University Park, PA 16802, USA}
\email{veresp@psu.edu, nnp@astro.psu.edu}
\def\ve{\varepsilon}
\def\gbm{{\it GBM }}
\def\lat{{\it LAT }}
\def\fermi{{\it Fermi }}

\newcommand{\Mesz}{{M\'esz\'aros}}

\def\mathnew{\mathsurround=0pt}
\def\simov#1#2{\lower .5pt\vbox{\baselineskip0pt \lineskip-.5pt
       \ialign{$\mathnew#1\hfil##\hfil$\crcr#2\crcr\sim\crcr}}}

\def\siml{\mathrel{\mathpalette\simov <}}
\def\beq{\begin{equation}}
\def\enq{\end{equation}}
\def\bea{\begin{eqnarray}}
\def\ena{\end{eqnarray}}

\def\L54{L_{54}}
\def\E55{E_{55}}
\def\et3{\eta_3}
\def\th1{\theta_{-1}}
\def\r07{r_{0,7}}
\def\x05{x_{0.5}}
\def\et600{\eta_{600}}
\def\et3{\eta_3}

\def\fflunit{\hbox{~erg cm}^{-2}~\hbox{s}^{-1}}
\def\eps{\epsilon}
\def\ve{\varepsilon}
\def\cm{\hbox{~cm}}

\def\gev{\hbox{~GeV}}
\def\GeV{\hbox{~GeV}}
\def\MeV{\hbox{~MeV}}
\def\kev{\hbox{~keV}}
\def\keV{\hbox{~keV}}
\def\eV{\hbox{~eV}}

\begin{document}
%\tableofcontents

\begin{abstract}
Most Fermi GRB spectra appear as either a broken power law extending to GeV
energies or as a broken power with a separate GeV power law component.
Here we show that such spectra can be understood in terms of magnetically
dominated relativistic jets where a dissipative photosphere produces the
prompt MeV emission, which is extended into the GeV range by inverse Compton
scattering in the external shock, with possible contributions from a reverse
shock as well. The bulk Lorentz factors required in these models are in the
range of 300-600, and the MeV-GeV time delays arise naturally. In some cases
an optical flash and a sub-dominant thermal component are also present.
\end{abstract}

\section{Introduction}
\label{sec:intro}

The GRB spectra observed with the \lat detector on \fermi reveal a diversity of
appearances. Some have spectra consistent with a single canonical Band function
\citep{Band+93} extending to the highest detected multi-GeV energies, while in
others at the higher end of the Band spectrum a second, flatter spectral
component emerges, which sometimes appears to have a cutoff
\citep{Pelassa+11-latgrbrev,Peer11-fermigrb}.  The origin of such a dichotomy, as
well as the extent of the applicability of such spectra for deriving lower
limits on the bulk Lorentz factor \citep{Kocevski+12LATpaucity} are the subject
of ongoing research and discussions.

Here we investigate possible scenarios for these two spectral types in the
context of magnetically dominated, baryon loaded outflows, where the prompt MeV
emission arises in the jet photosphere. Magnetically dominated models have
recently received increased attention \citep{Drenkhahn+02, Lyutikov03grbmag,
Giannios+07photspec, Giannios11peak, Komissarov+09maggrb, McKinney+11switch,
Metzger+11magcr, Meszaros+11gevmag, Bosnjak+12delay, Zhang+11icmart}, while
photospheric emission, either in baryonic or magnetic models where dissipation
occurs near the photosphere, have been shown to be able to produce Band-like
non-thermal spectra \citep{Peer+06phot, Lazzati+11phot, Beloborodov10phot,
Ryde+10phot090902, Peer+12thermal}.  In the scenarios that we discuss here we
combine these two elements and add a third one, assuming that the prompt
spectrum is a combination of the emission from two distinct regions, namely, a
magnetically dominated dissipative photosphere and the external shock. We argue
that the photosphere is initially responsible for the low energy (\gbm) part of
the Band-like prompt spectrum, the final observed spectrum being modified by
inverse Compton scattering in the external shock, which results in a delayed
high energy (\lat) component. The latter includes a forward and possibly also a
reverse shock contribution, dominated by self-inverse Compton (SSC) as well as
external inverse Compton (EIC) of the photospheric photons. We consider here
mainly the radial structure of the outflow, and restrict ourselves to leptonic
mechanisms.

Our primary goal here is to explore the generic features of such models, and to
test the extent to which they qualitatively lend themselves to the
interpretation of \lat spectra. That is, we do not perform detailed fits to
individual bursts, which at this point would be premature given the preliminary
nature and the uncertainties in the models. The purpose is to explore the
connection between elements of the physical model and the presence or absence
of broad spectral features, and how these depend on typical burst parameters.
We concentrate on the prompt emission, from the time of the trigger up to times
comparable to the deceleration time, when the external forward shock (and the
reverse, if present) have fully developed. The latter should be representative
of the physics in the late afterglow, which we do not address specifically
here.  We then qualitatively compare our results to several of the archetypal
\lat bursts, such as GRB 090902B \citep{Abdo+09-090902B}, which has a Band
component and an extra	high energy power law (PL) second component; GRB
090926, which also has an extra spectral component and shows a high energy
cutoff \citep{Ackermann+11-090926}; and GRB 080916C, which shows a single Band
component extending to multi-GeV \citep{Fermi+09-080916-sci}.

We show that in the context of these models, this variety of appearances can be
attributed to differences in the strength of the various radiation components
under plausible variations of the input physical parameters. Previous
investigations	\citep[e.g.,][]{Toma+11phot, Fraija+12ssc} have considered a
connection between the extra high energy component and inverse Compton and/or a
reverse shock in a baryon dominated outflow. Here the magnetically dominated
dynamics, photospheric input spectrum and consideration of separate SSC and EIC
components from both forward and reverse shock result in a different and
broader range of possible outcomes.  Depending on the parameters, the SSC or
EIC from the forward shock and the reverse shock (if the latter is present) can
either result in an extra high energy component, or sometimes in an apparent
single Band component extending to the highest energies.  We show that, even if
there is a pair-production cutoff in the nonthermal prompt emission at a few or
a few tens of MeV, the external inverse Compton radiation from the reverse
shock can constitute a natural extension of the absorbed emission to form a
continuous Band spectrum out to higher energies. We argue that if a reverse
shock does not develop, the dominant high-energy components will be the forward
shock SSC or EIC, resulting in a distinct extra component. For some parameters,
however, the latter can also be produced by a reverse shock.  A further result
from these calculations is that, since the high energy component arises mainly
in the external shock, where the compactness parameter is low, these high
energy photons	do not constrain via $\gamma\gamma$ cutoff considerations the
bulk Lorentz factor to as high values as considered in many of the recent
analysis papers.

In \S \ref{sec:magmod} we discuss the basic magnetic jet model used. In \S
\ref{sec:phot} we give the details of the dissipative photospheric (prompt)
spectrum. In \S \ref{sec:ext} we discuss the various external shock synchrotron
and inverse Compton radiation components.  In \S \ref{sec:results} we present
numerical results for various cases.  In \S \ref{sec:disc} we discuss our
results and conclusions.

\section{Magnetic Model}
\label{sec:magmod}

We assume a total luminosity $L_{t}=10^{53} L_{t,53}$ erg/s is released at a
radius $r_0=10^7 r_{0,7} $ cm, lasting for $10$ s in the central engine frame.
This is initially (at $r_0$) mainly magnetic, which at larger radii gets
gradually converted into bulk kinetic energy and non-thermal radiation. We
assume that at the dissipative photosphere a fraction $\zeta_{r}$ of $L_t$ is
released as prompt radiation, whose spectrum is discussed in the next section.
A nominal value for this here is $\zeta_r \simeq 0.5$, which we assume to
consist mainly of a non-thermal component $\zeta_{nt} \sim \zeta_r$ giving the
Band power law (PL) spectrum (a smaller fraction of $\zeta_r$ is in a thermal
component $\zeta_{th} < \zeta_{nt}$, discussed in \S \ref{sec:th}, so that
$\zeta_r=\zeta_{nt}+\zeta_{th}$). The remaining energy fraction beyond the
photosphere $\zeta_k$, after the Lorentz factor reaches its saturation
(coasting) value, is in kinetic energy form.  About half of this, $\zeta_{d}
\simeq 0.25$, is assumed to be radiated away in the external shock (forward and
reverse) when deceleration starts, while the other half, $\zeta_{ag} \simeq
0.25$, continues as kinetic energy of the decelerating ejecta, and is
eventually radiated away in the longer afterglow phase. Thus, $\zeta_{r}+
\zeta_{k}=1$. Most of our results are evaluated for $\zeta_{r}=1-
\zeta_{k}=0.5$, and we use these nominal numerical fractions unless stated
otherwise. These factors are used to define the respective luminosities, e.g.,
$L_{k,53}=(1-\zeta_{r}) L_{t,53}$, etc., and we use the $Q=10^x Q_x$ convention
in cgs units, except, e.g. for the coasting Lorentz factor
($\eta=600~\eta_{600}$).

We consider a continuous outflow in which the central engine is active for
$t_{\rm 0,obs}= 20 ~t_{1.3} (1+z)/2$ s, while the deceleration time is $\sim$ few
seconds resulting in a modestly thick-shell deceleration region, which for
simplicity will be approximated as a thin shell. We assume a constant density
external medium ($n=1 \ n_0 \cm^{-3}$) and a burst with a fiducial redshift and
luminosity distance of $z=1$ or $D_L=2\times10^{28} \  D_{28.3}\cm $.

In magnetically dominated models the magnetic fields become rapidly transverse,
in a striped magnetic structure \citep{1984ApJ...283..710K}.  The Lorentz
factor in such jets will increase more slowly than in a matter dominated case,
e.g. after a few characteristic lengths above $r_0$ the average Lorentz factor
will increase as $r^{m}$, where $m$ depends on the geometry assumed at the
lower boundary \citep[e.g.,][]{McKinney+11switch,Narayan+10numjet}. When
magnetic reconnection plays a role, the expected Lorentz factor behavior is
$\gamma\propto r^{1/3}$ until it saturates
\citep{Drenkhahn02, Giannios+07photspec, Metzger+11grbmag,Meszaros+11gevmag}, which
we assume in what follows.

\begin{equation}
{\Gamma(r)}\propto \left\{
\begin{array}{lll}
 r^{1/3}	&	{\rm if}	& r<r_{sat}\\
 {\rm const.}&	{\rm if}	& r_{sat}<r<r_{dec},
\end{array}
\right.
\label{eq:accel}
\end{equation}
where $r_{sat}$ is the saturation radius where $\Gamma \to \eta \equiv
L_{t}/{\dot M}c^2$ and	$r_{dec} > r_{sat}$ is the deceleration radius. We will
concern ourselves here only with the behavior before and in the neighborhood of
the deceleration radius, and do not consider the longer term afterglow phase.
A feature worth stressing is that an acceleration behavior such as
equ. (\ref{eq:accel}) implies a comoving volume element $V'\propto r^2\Gamma(r)$
scaling leading to a radius dependence of the comoving particle density $n'$,
temperature $T'$, etc.	which differs from that in the usual baryon-dominated
$\gamma\propto r$. Also, the saturation radius occurs at significantly larger
radii than in the matter dominated acceleration cases, if other relevant
quantities are the same. Furthermore, the scattering photospheric radius (see
equ. [\ref{eq:rphot}]) occurs generally before the saturation radius.

\section{Prompt Photospheric Radiation}
\label{sec:phot}

The photospheric radius is obtained by setting the scattering optical depth to
unity for the ejecta, i.e. $\tau_T = L_{r}/(4\pi r^2 m_p c^3 \eta \Gamma) \sigma_T
r/(2\Gamma) = 1$, if the $e^+e^-$ pair contribution can be ignored.  The effect
of pairs will be that they will increase the photospheric radius.  With the
increased radius variables have to be recomputed, resulting in a different
cutoff energy and in a different number of pairs. At the end of this iterative
process, we find that the photospheric radius increases by a factor of $\sim 5$
for the nominal parameters  \citep[see also ][]{Bosnjak+12delay}.  We address the
details of pair creation at the end of this section.

For $\Gamma=(r/r_0)^{1/3}$ the photospheric radius at which $\tau_T=1$ is
\begin{equation}
r_{ph} =\left(\frac{L_r r_0^{2/3} \sigma_T}{8\pi m_p c^3 \eta}\right)^{3/5}=
6.5\times 10^{12} \ L_{t,53}^{3/5} \zeta_r^{3/5} r_{0,7}^{2/5} \eta_{600}^{-3/5} \cm
\label{eq:rphot}
\end{equation}
for a nominal $\eta=600$, where $\sigma_T$ is the Thomson cross section. The
photospheric radius occurs in the accelerating phase for any realistic set
of parameters, for the magnetized dynamics described in the previous section.
The saturation radius, where the acceleration ceases and the ejecta starts to
coast with $\Gamma \sim \eta \simeq$ constant is, for a magnetically dominated
jet, at
\begin{equation}
r_{sat} = r_0 \eta^3= 2.2\times10^{15}~ r_{0,7} \eta_{600}^3 \cm.
\label{eq:rsat}
\end{equation}
While compared to baryonic outflow models the photosphere is relatively farther
away from the central engine, both the photosphere and the saturation radius
remain generally smaller than the deceleration radius,
\begin{equation}
r_{dec}=\left(\frac{3 L_k t_0}{4 \pi n m_p c^2 \Gamma^2 }\right)^{1/3} = 4.8
\times 10^{16}	L_{t,53}^{1/3} (1-\zeta_r)^{1/3} t_{1.3}^{1/3} n_0^{-1/3}
\eta_{600}^{-2/3} \cm.
\label{eq:rdec}
\end{equation}
This is the radius at which we calculate, in \S \ref{sec:ext}, the interaction
of the photospheric photons with the external shock electrons, resulting in SSC
and EIC components. Using the scaling relations of the magnetic dynamics, the
density decreases with radius as $n'_b\propto V'^{-1} \propto r^{-7/3}$ for
$r<r_{sat}=r_0 \eta^3$ and $n'_b\propto r^{-2}$ for $r_{sat}<r$, that is,
$n'_b(r) = n'_{0} (r/r_{0})^{-7/3}$ for $r<r_{sat}$ and $n'_b(r) = n'_{sat}
(r/r_{sat})^{-2} $ for $ r_{sat}< r$.  Thus, the comoving baryon density at the
deceleration radius is $n'_{b,dec}= 5.3 \times 10^4~
(1-\zeta_{r})^{-2/3}L_{t,53}^{1/3}\eta_{600}^{-2/3} n_0^{2/3}t_{1.3}^{-2/3}~
\cm^{-3}$.

\subsection{Nonthermal Photospheric Spectrum}
\label{sec:nonth}

%ref kerdes 1
Near the scattering photosphere, a prompt spectrum with a high radiation
efficiency can be produced through dissipative effects. For non-magnetic
outflows, such photospheres were considered by, e.g. \citet{Rees+05photdis,
Peer+06phot, Beloborodov10phot}, while for magnetic photospheres dissipation of
magnetic energy via reconnection, turbulence dissipation or the associated
semi-relativistic shocks was considered by, e.g. \citet{Giannios+07photspec,
Giannios11peak, McKinney+11switch, Meszaros+11gevmag}.	This generally can
convert  some fraction $\zeta_r$ of the initial jet energy $L_t$ into a
prompt burst of radiation, the rest remaining available as kinetic energy of
the coasting ejecta beyond the saturation radius.
%ref kerdes1 folyt
The spectrum of an unmagnetized dissipative  outflow in the photospheric
neighborhood has been calculated by \cite{Beloborodov10phot}, who performed a
radiative transfer calculation including both thermal and non-thermal electrons
from nuclear collisions which results in a Band spectrum with spectral indices
in the range of those observed (see also \cite{Vurm+11phot} for inclusion of
magnetic fields up to $\epsilon_B\sim 1$). Magnetic photospheres where
reconnection leads to magnetic turbulence and/or dissipation combined with
Comptonization effects involving purely thermal electrons were calculated by
\cite{Thompson94} and \cite{Giannios+07photspec}, resulting also in Band-like
spectra with spectral slopes close to the average observed values $\alpha=1$
and $\beta=2$.	Of course, the dissipation of energy can occur in magnetic
cases both below and above the photosphere, up to the saturation radius
\citep{Giannios08prompt}, and also in non-magnetic cases scattering occurring
both below and for some distance above the photosphere can affect the spectrum
\citep{Peer+06phot, Beloborodov10phot}.  These calculations indicate that the
spectral peak occurs naturally in the neighborhood of $\ve_{pk}\sim 300$ keV.
Dissipation radii larger than the photosphere would permit also multi-GeV
prompt photons to escape if these are indeed produced as shown by
\citet{Giannios08prompt, Beloborodov10phot, Vurm+11phot}.

%ref kerdes 3
Here we do not specialize to a specific dissipation mechanism, and we consider
an idealized photospheric model where some fraction $\zeta_r\sim 0.5$ of the
initial outflow luminosity is converted into prompt radiation near the nominal
radius $r_{ph}$ of the photosphere, producing a Band-like spectrum whose slopes
are in the range of the observed values. In order to connect this spectrum to
the magnetic properties of the flow, we assume that the spectral peak is given
by the synchrotron spectral peak of the minimum energy electrons accelerated in
the dissipative photosphere, which as it happens turns out also to be in the
observed range of $\sim 300$ keV. (Note that, in principle, the prompt emission
model could also be taken to be a nonmagnetic photosphere, as in
\cite{Peer+06phot, Beloborodov10phot}; the photospheric radius would be
different, but the upscattering in the external shock would differ only insofar
as caused by any possible differences in the photospheric seed spectrum. Also,
the range of radii over which the prompt spectrum forms does not have an impact
on the upscattered spectrum, as long as the external shock lies significantly
beyond the dissipation and saturation radii).

The value of the Lorentz factor at the magnetic photosphere is
\begin{equation}
\Gamma_{ph} = (r_{ph}/r_0)^{1/3} = 87 \ L_{t,53}^{1/5}\zeta_r^{1/5} r_{0,7}^{-1/5}
\eta_{600}^{-1/5}.
\label{eq:Gammaph}
\end{equation}
At this photosphere, the magnetic dissipation and/or collisional effects such
as $n,p$ decoupling will lead to a significant non-thermal electron component.
E.g. for reconnection acceleration or semi-relativistic shocks associated with
reconnection \citep{Meszaros+11gevmag}, a typical minimum comoving electron
random Lorentz factor $\gamma_{e,ph}\sim 600 \Gamma_r$, coincidentally of
similar order of magnitude as $\eta$.
The magnetic field at the photosphere is $B_{ph}' = (32 \pi \epsilon_B m_p c^2
{n'}_b)^{1/2} \Gamma_{r}\  =\  1.7 \times 10^6 \ L_{t,53}^{-1/5}
(1-\zeta_r)^{1/2}\zeta_r^{-7/10} r_{0,7}^{-3/10}  \eta_{600}^{1/5}
\epsilon_{B,0}^{1/2}  \Gamma_r $ G.
Here $n'_b$ is the comoving baryon density evaluated at the photosphere,
$n'_{b,ph}=L_{k}/(4\pi m_p c^3 r_{ph}^2 \eta \Gamma_{ph})=2.0\times 10^{13}
~L_{t,53}^{-2/5} \zeta_r^{-7/5} (1-\zeta_r) r_{0,7}^{-3/5} \eta_{600}^{2/5} ~
{\rm \ cm}^{-3}$, Here as before, $L_k$ and $~L_r$ are expressed as functions
of $L_t$ and $\zeta_r$.  The observed synchrotron spectral peak of the
photosphere will	be
\begin{equation}
{\varepsilon}_{sy}^{ph,obs} = \varepsilon_{br} = \frac{3 e h B'_{ph}}{4 \pi m_e
c} \gamma_{e,ph}^2 \frac{\Gamma_{ph}}{1+z}= 310 ~\zeta_r^{-1/2}
(1-\zeta_r)^{1/2}	r_{0,7}^{1/2} \epsilon_{B,0}^{1/2}\Gamma_r^3
\left(\frac{1+z}{2}\right)^{-1} \kev, \label{eq:ephobs} \end{equation}
corresponding to the Band peak.  The magnetic field energy fraction at the
photosphere is still $\epsilon_B \sim 1$, but dissipation will lead to values
$\epsilon_{B,FS} \lesssim 0.1$ by the time the flow reaches the deceleration
radius.
Following \cite{Thompson94,Giannios+07photspec} and \cite{Beloborodov10phot},
we assume that the emergent spectral shape is of the Band form, taking for its
peak value dependence on the flow parameters the expression (\ref{eq:ephobs}).
As a nominal photon number low energy spectral index we take $\alpha \simeq 1$
(where $N_\ve(\ve)\propto \ve^{-\alpha}$) and for the high energy spectral index
$\beta \simeq 2.4$.  The flux in energy per energy units is then
\begin{equation}
F_\varepsilon= A \left\{
\begin{array}{ll}
 \left(\frac{\varepsilon}{\varepsilon_{br}}\right)^{-\alpha+1}
& \textrm{if}\, \varepsilon \leq \varepsilon_{br}
\\
 \left(\frac{\varepsilon}{\varepsilon_{br}}\right)^{-\beta+1}
& \textrm{if}\, \varepsilon > \varepsilon_{br}
\end{array}
\label{eq:Band}
\right.
\end{equation}
The flux is related to the luminosity via $L/4\pi D_L^2 = \int F_\epsilon
d\epsilon$ which yields the normalization factor $A=L_{r}/(4\pi D_L^2
\varepsilon_{sy}^{ph,obs} (1/(2-\alpha) + 1/(\beta-2)))= 9.1 \times 10^{-9}~
L_{t,53} \zeta_r^{3/2} (1-\zeta_r)^{-1/2} D_{L,28.3}^{-2}  r_{0,7}^{1/2}
\epsilon_{B,0}^{-1/2} \Gamma_r^{-3} ((1+z)/2) \fflunit {\keV}^{-1} $.  The
$\varepsilon F_\varepsilon$ peak of the nonthermal spectrum will arise at
$\varepsilon_{br}$ and at $(\varepsilon F_\varepsilon)^{\max}=\varepsilon_{br}
A=2.8\times10^{-6}~ L_{t,53}\zeta_r  D_{L,28.3}^{-2} ~\fflunit$.	At low
energies a self-absorption break is expected. In our case, this will occur at $
\varepsilon_{SA} \approx 6.8 ~ L_{t,53}^{1/5}\zeta_r^{-1/5}(1-\zeta_r)^{1/5}
\eta_{600}^{7/15} \epsilon_{B,0}^{1/3} \keV$ \citep{Guetta+03highE}.

The high energy branch of the photospheric spectrum (\ref{eq:Band}) extends up
to an energy which is model dependent. We discuss here two scenarios.  In the
first scenario, photons above the photospheric spectral peak (\ref{eq:ephobs})
are upscattered as a result of interactions with electrons associated with
magnetic turbulent waves, up to an energy  $\ve' \approx m_e c^2$ in the jet
frame \citep{Thompson94}.  In our case, this corresponds to an observer frame
cutoff at
\begin{equation}
\ve_{h}\approx (4/3) m_e c^2 \Gamma_{ph}/(1+z)\approx 30 ~L_{t,53}^{1/5}\zeta_r^{1/5}
r_{0,7}^{-1/5} \eta_{600}^{-1/5} \left(\frac{1+z}{2}\right)^{-1} \MeV .
\label{eq:ephthom}
\end{equation}
This corresponds to  $\ve'\sim m_e c^2$ in the comoving frame, and is below the
$\gamma\gamma$ cut-off expected from pair production against photospheric lower
energy photons (see below), so no pairs are created.

In the second scenario, we assume that the magnetic reconnection regions or the
shocks associated with them are coherent over long enough times that electrons
can be accelerated, e.g. via a Fermi mechanism, to a power law extending above
the previous $\ve_h$ to photon energies  sufficiently high to be subject to
$\gamma\gamma$ interactions.  We estimate the $\gamma\gamma$ annihilation
energy from requiring the pair optical depth to be unity against target photons
of energy $\varepsilon_{at}$.
\begin{equation}
\varepsilon_{\gamma\gamma}=(\Gamma_{ph} m_e c^2)^2/[(1+z)^2\varepsilon_{at}],
\label{eq:ephgg}
\end{equation}
where $\tau_{\gamma \gamma} \sim 11/180 \sigma_T N_{>\varepsilon_{at}}/4\pi
r_{ph}^2 $ \citep{Lithwick+01gev,Murase+08pair}, where $N_{>\varepsilon}$ is
the number of photons with energies higher than $\ve$. This will result in an
observer-frame cut-off energy $\varepsilon_{\gamma\gamma}\sim 30-100$ MeV. If
$\ve_{\gamma \gamma}>\ve_{at}$ is not satisfied, the cutoff will be at $\sim
\Gamma_{ph} m_e c^2/(1+z)$. Above the cut-off, the spectrum becomes steeper,
the exact post cut-off spectral index depending on the spectrum and photon
spatial distribution.  A simple slab approximation, as discussed for GRB
090926A \citep{Ackermann+11-090926}, results in a steepening of the high energy
slope index by $\beta-1$ above the $\varepsilon_{\gamma\gamma}$ energy
\citep[see also ][]{Beloborodov10phot}.

These pairs also radiate in the magnetic field of the prompt emission site, and
this will result in a low energy synchrotron component peaking at tens of
electron volts.  As we discuss in \S \ref{sec:disc}, this component is a good
candidate for the bright prompt optical emission observed in some bursts.  The
pair synchrotron component will be in the fast cooling regime, with
$\gamma_{\pm,m} =\ve_{\gamma\gamma} (1+z)/(2 \Gamma_{ph} m_e c^2)=3.9$ and
$\ve_{\pm,{\rm peak}}\approx3.2 \eV$ \citep[e.g.][]{Toma+11pop3}. The
functional dependence is different depending on the different cases for
$\ve_{\gamma\gamma}$.  The peak of the $\ve F_\ve$ spectrum is $7.7\times
10^{-9} \fflunit$  for nominal parameters.

\subsection{Thermal Component of the Photospheric Spectrum}
\label{sec:th}

In the presence of collisional or magnetic dissipation, in addition to a
nonthermal component one expects also a thermal component, whose luminosity
$L_{th}=\zeta_{th}L_t$ should have a quasi-blackbody spectrum. In our case,
this peaks in the soft X-rays \citep{Meszaros+11gevmag}, due to the different
temperature scaling with radius for the magnetic dynamics, comprising a
fraction $\zeta_{th}$ of the luminosity of the photosphere.  This thermal
component can be calculated from the initial $T_0=\left(\frac{L_{t}}{4\pi r_0^2
a c \Gamma_r^2}\right)^{1/4} =2.1 ~L_{t,53}^{1/4} r_{0,7}^{-1/2}
\Gamma_r^{-1/2}~\MeV$ at the initial radius $r_0 =10^7 r_{0,7} {\cm}$.

Above $r_0$ the magnetically dominated jet dynamics $\Gamma\propto r^{1/3}$
implies a comoving volume is $V'\propto r^2\Gamma \propto r^{7/3}$ in the
acceleration regime.  Thus, the  temperature will decrease more gradually with radius
than in the baryon-dominated case (where $T'\propto r^{-1}$), as $T'\propto
\rho'^{\hat{\gamma}-1}\propto\rho'^{1/3}\propto V'^{-1/3}\propto r^{-7/9}$, or
$T'(r)=T_0 (r/r_0)^{-7/9}$, where $\hat{\gamma}=4/3$ is the adiabatic exponent
for a relativistic gas. Thus, at the photosphere $T(r_{ph}) = 2.7 ~
L_{t,53}^{-1/60} \zeta_r^{-4/15} \eta_{600}^{4/15} r_{0,7}^{-7/30}
\Gamma_{r}^{-1/2} \left(\frac{1+z}{2}\right)^{-1} ~\kev$.

The corresponding thermal luminosity is $L_{th}=6.5\times 10^{49}$ erg/s, which
is low for producing the prompt emission, as well as being too soft. This
blackbody component from the photosphere  thus peaks in the soft X-rays at
$(\ve F_{\ve,BB})^{{\rm peak}}\approx 9.5\times10^{-9}~L_{t,53}^{11/15}
\zeta_r^{-4/15} \eta_{600}^{4/15} r_{0,7}^{4/15} \fflunit$. This thermal
component is similar to the one found by \citet{Page+11Xthermal}.

\section{External Shock Radiation Spectral Components}
\label{sec:ext}

\subsection{Forward Shock (FS) Synchrotron}
\label{sec:fs}

The forward shock develops at the deceleration radius, where the jet has plowed
up an amount of external mass roughly equal to $1/\eta$ times of the ejecta
mass.  The electrons in the shock will be accelerated into a relativistic
energy distribution, and will undergo cooling  through synchrotron emission and
by inverse Compton scattering off external (photospheric) and their own
(synchrotron) photons. At this deceleration radius the Lorentz factor has
roughly halved from its coasting value.  The time of the deceleration is
$t_{dec}=r_{dec}/(2\eta^2 c) (1+z) \approx4.4~ L_{t,53}^{1/3} (1-\zeta_r)^{1/3}
t_{1.3}^{1/3} n_0^{-1/3} \eta_{600}^{-8/3} (1+z)/2~ {\rm s}$.

While at the photosphere the magnetic field parameter $\epsilon_{B}^{ph}$ is
close to unity for a magnetically dominated model, after  magnetic dissipation
ceases one expects this parameter to be much less, the outflow becoming essentially
baryon dominated \citep[e.g.,][]{Zhang+11icmart}. Here we assume that at the
deceleration radius $\epsilon_{B}^{FS} \lesssim 0.1$.

The magnetic field in the forward shock is then
$B'_{FS} = (32 \pi m_p c^2 \epsilon_B \Gamma_{FS}^2 n)^{1/2} \approx 74~
 \eta_{600} \epsilon_{B,-1}^{1/2} n_0^{1/2} ~\rm{G}$.
The cooling Lorentz factor is $\gamma_c=6 \pi m_e c/((1+Y) \sigma_T B_{FS}^2
\Gamma t_{dec}) = 3/(8 (1+Y)\sigma_T (m_p/m_e) \epsilon_B n r_{dec} \eta)=98~
(1-\zeta_r)^{-1/3} L_{t,53}^{-1/3} t_{1.3}^{-1/3} \epsilon_{B,-1}^{-1} n_0^{-2/3}
\eta_{600}^{-1/3} (1+Y)^{-1}$, where $Y = (- 1 + \sqrt{1 + 4\epsilon_e/
\epsilon_{B,FS}})/2\approx0.1$	is the Compton parameter.
The minimal (injection) Lorentz factor is \begin{equation}
\gamma_m=\epsilon_e\frac{m_p}{m_e}\frac{p-2}{p-1} \Gamma \equiv 3100 \
\epsilon_{e,-2} \eta_{600} g_{p,2.4}, \label{eq:gammam} \end{equation} where
$g_{p,2.4}$ is the fraction $(p-2)/(p-1)$ normalized to $p=2.4$.  Thus, the
electrons responsible for the synchrotron radiation are in the fast cooling
regime, and their distribution is given by $dN_e/d\gamma\propto\gamma^{-2}$ if
$\gamma_c \leq \gamma < \gamma_m$ and $dN_e/d\gamma\propto\gamma^{-p-1}$ if
$\gamma \geq \gamma_m$.

The observer frame energy of photons radiated by electrons with a random Lorentz
factor $\gamma_m$ in the forward shock is
\begin{equation}
{\varepsilon}_m=\frac{3 h e B'_{FS}}{4 \pi m_e c}\gamma _m^2 \frac{\Gamma_{FS}}{1+z}=
3.8 \  \epsilon_{B,-1}^{1/2} \epsilon_{e,-2}^2 n_0^{1/2}
\eta_{600}^4 g_{p,2.4}^2 \left(\frac{1+z}{2}\right)^{-1} \kev,
\end{equation}
while for electrons with a random Lorentz factor $\gamma_c$ the observed photon
energy is
\begin{equation}
{\ve}_c= 3.7  ~L_{t,53}^{-2/3}(1-\zeta_r)^{-2/3}
t_{1.3}^{-2/3} \epsilon_{B,-1}^{-3/2} n_0^{-5/6} \eta_{600}^{4/3} (1+Y)^{-2}
\left(\frac{1+z}{2}\right)^{-1}~\eV.
\end{equation}
The peak flux density of the FS synchrotron spectrum occurs, for fast cooling, at
$\varepsilon_c$, and is given by \citep[e.g. ][]{1998ApJ...497L..17S}:
\begin{equation}
F_{max}^{FS}(\varepsilon_c)=
\frac{4 \pi r_{dec}^3 n}{3} \frac{m_e c^2 \sigma_T \eta B'_{FS}}{12 \pi q_e
D_L^2}=0.15 ~L_{t,53} (1-\zeta_r) t_{1.3} \epsilon_{B,-1}^{1/2}
n_0^{1/2} D_{L,28.3}^{-2} ~{\rm Jy}.
\end{equation}
The maximum Lorentz factor attainable by the electrons is calculated by
equating the acceleration timescale in the shock to the radiation timescale,
which gives $\gamma_M=(3 q_e/g_M \sigma_T B_{FS}')^{1/2}\sim 5.4\times10^6~
\eta_{600}^{-1/2} \epsilon_{B,-1}^{-1/4} n_0^{-1/4}$, where $g_M$ is a
numerical factor of order unity. The corresponding photon energy is
$\varepsilon_{FS,M}=3 h q_e B_{FS}'/(4\pi m_e c) \gamma_M^2 \eta/(1+z) =11.2~
\eta_{600} ((1+z)/2)^{-1}$ GeV.

The scattering optical depth in the FS is given by
\begin{equation}
\tau_{FS}=\frac{N_e \sigma_T}{4 \pi R^2}=\frac{4\pi r_{dec}^3 n
\sigma_T}{3\times4\pi r_{dec}^2}=1.1\times 10^{-8}~
L_{t,53}^{1/3} (1-\zeta_r)^{1/3} t_{1.3}^{1/3} \eta_{600}^{-2/3} n_0^{2/3}
\end{equation}
where $N_e$ is the number of electrons in the forward shock.

\subsection{Reverse Shock (RS) Synchrotron }
\label{eq:rs}

The reverse shock, if it develops, becomes strongest at the deceleration
radius.  A contact discontinuity (CD) separates the FS from the RS, and in the
frame of the CD the RS travels backwards.  The pressure is the same in both
sides of the CD, and consequently the magnetic energy density will be the same
as well, under the usual assumption that the magnetic field is turbulently
generated and $\eps_B$ is taken to be the same in both sides.

The characteristic frequencies for the RS will be: $\varepsilon_{m}^{RS} =
\varepsilon_{m}^{FS}/ \Gamma^2=1.1\times10^{-5}  ~ \epsilon_{B,-1}^{1/2}
\epsilon_{e,-2}^2  n_0^{1/2} \eta_{600}^2  g_{p,2.4}^2
\left(\frac{1+z}{2}\right)^{-1} \keV $ and $\varepsilon_{c}^{RS} =
\varepsilon_{c}^{FS}$ \citep[e.g. ][]{Sari+00refresh}.	The RS cooling
frequency can differ by about $10\%$, but we will ignore this difference.
Generally the RS peak flux will be a factor $\Gamma$ higher than in the FS. In
our case the radiative regime changes from the FS to RS: while the FS electrons
are in fast cooling, the reverse shock is in the slow cooling case. We
calculate the peak flux $F_{\nu,max}^{RS}$  from
\begin{equation}
F_{max}^{RS}(\varepsilon_m)=\frac{N_e P_{max} (\nu)}{4 \pi D_L^2}=\frac{L_k
t_0}{\eta m_p c^2} \frac{m_e c^2 \sigma_T \eta B}{3 q_e} \frac{1}{4 \pi
D_L^2}= 92 ~ L_{t,53}(1-\zeta_r)  t_{1.3} \eta_{600} \epsilon_{B,-1}^{1/2}
n_0^{1/2}D_{L,28.3}^{-2}~ {\rm Jy}.
\end{equation}
It is unclear, in an initially magnetically dominated outflow, whether the RS
will develop or not, depending on various assumptions \citep{Zhang+05rsmag,
Giannios+08rsmag, Zhang+11icmart,Narayan+11magn}. In the simplest cases the
Alfv\'enic $\gamma_A'\sim \sqrt{1+\sigma}$ Lorentz factor will be high and
magnetic waves can carry information from the CD to the start of the ejecta,
suppressing the RS. Here $\sigma\approx F_{\ve,r}/F_{\ve,k}$ is the
magnetization parameter giving the ratio of the Poynting (magnetic) flux to the
kinetic flux. However, the result depends on the amount of dissipated magnetic
energy at radii before reaching the deceleration radius. If there is
significant portion of magnetic energy compared to the kinetic energy there may
be no reverse shock, while in the contrary case reverse shocks may form.

The optical depth of the RS region is important for calculating the flux of the
inverse Compton radiation.  We calculate the optical depth of the RS at
$r_{dec}$ from
\begin{equation}
\tau_{RS}=\frac{N_e \sigma_T}{4 \pi R^2}=\frac{L_k t_0 \sigma_T}{\eta m_p c^2
4\pi r_{dec}^2}=6.4\times 10^{-6} L_{t,53}^{1/3}(1-\zeta_r)^{1/3}
t_{1.3}^{1/3}\eta_{600}^{1/3} n_{0}^{2/3}
\end{equation}
where $N_e$ is the number of electrons in the reverse shock. Note that at
$r_{dec}$ the optical depth of the RS  is $\eta$ times the optical depth of the
FS as expected.

%\section{External Shock Inverse Compton Components}
%\label{sec:ic}

\subsection{Forward Shock Self-Compton (FS-SSC)}
\label{sec:fs-ssc}

At the deceleration radius the forward shock (and the reverse shock, if
present) accelerated electrons will cool also by inverse Compton interactions
with their own synchrotron photons.  The forward shock at the deceleration is
in the synchrotron fast cooling regime, for the parameters considered here, and
the Compton parameter is $Y= (-1+\sqrt{1+4\eta_{SSC} \epsilon_e/ \epsilon_B})/2
\approx \epsilon_e/\epsilon_B=0.1$, where $\eta_{SSC}=\min
((\gamma_c/\gamma_m)^{2-p},1)$ and $\eta_{SSC}=1$ is valid for the fast cooling
case.

The SSC component of the FS will also be in the fast cooling radiative regime,
and the minimal and the cooling energies will be given by $\varepsilon_m^{SSC}
\approx 2 \gamma_m^2 \varepsilon_m = 75  ~\epsilon_{B,-1}^{1/2}
\epsilon_{e,-2}^4 n_0^{1/2} \eta_{600}^6 g_{2.4}^4  ((1+z)/2)^{-1} \gev $ and $
\varepsilon_c^{SSC} \approx 2 \gamma_c^2 \varepsilon_c = 70 ~L_{t,53}^{-4/3}
(1-\zeta_r)^{-4/3} t_{1.3}^{-4/3} \epsilon_{B,-1}^{-7/2} n_0^{-13/6}
\eta_{600}^{2/3} (1+Y)^{-4} ((1+z)/2)^{-1} \keV$ respectively.

The amplitude of the FS SSC component is determined from: $(\ve^{SSC}
F_{\ve})^{SSC}_{\rm peak} =	Y \varepsilon^{FS}_m F_{\varepsilon_m}^{FS}=
3.7\times 10^{-9} L_{t,53}^{2/3} (1-\zeta_r)^{2/3} t_{1.3}^{2/3}
\eta_{600}^{8/3} n_{1/3} \epsilon_{e,-2}^1 g_{p,2.4}^1	D_{L,28.3}^{-2}
\fflunit.$ The higher energy electrons can reach the Klein-Nishina (KN) regime.
This results in a break in the spectrum at $\varepsilon_{KN,c}^{SSC}=\gamma_c
m_e c^2 \eta/(1+z)=15 ~  L_{t,53}^{-1/3} (1-\zeta_r)^{-1/3}  t_{1.3}^{-1/3}
\epsilon_{B,-1}^{-1} n_0^{-1/3} \eta_{600}^{1/3}  (1+Y_{SSC})^{-1} \gev$.
Details of the KN break energy are in Appendix \ref{KN}.

\subsection{EIC scattering of photospheric nonthermal photons on external forward
shock electrons (FS-EIC)}
\label{sec:fs-eic}

The electrons in the forward shock will also lose energy by external inverse
Compton (EIC) as they upscatter the photons from the prompt emission
(photospheric) region.
 This effect was discussed by \cite{Beloborodov05gev} for interaction of the
prompt photons with both FS and RS. It was further proposed that a delay of the
order observed by Fermi arises within this setup
(see also \cite{Meszaros+94gev}).
The minimum and cooling frequencies of this component
will be $\varepsilon_m^{EIC} \approx 2 \gamma_m^2 \varepsilon_{br} = 6.2~
\zeta_r^{-1/2} (1-\zeta_r)^{1/2} \epsilon_{B,0}^{1/2}  \epsilon_{e,-2}^2
\eta_{600}^2 g_{p,2.4}^2 r_{0,7}^{-1/2} \Gamma_r^3 ((1+z)/2)^{-1} ~{\rm TeV}$
and $ \varepsilon_c^{EIC} \approx 2\gamma_c^2 \varepsilon_{br} = 6.0~
L_{t,53}^{-2/3} \zeta_r^{-1/2}(1-\zeta_r)^{-1/6} t_{1.3}^{-2/3}
\epsilon_{B,-1}^{-2} \epsilon_{B,0}^{1/2} n_0^{-4/3} \eta_{600}^{-2/3}
(1+Y)^{-2}r_{0,7}^{-1/2} \Gamma_r^3 ((1+z)/2)^{-1} \gev$, {where we
differentiate between prompt and FS magnetic parameter by using
$\epsilon_{B,0}$ and $\epsilon_{B,-1}$ respectively.}

In the absence of KN suppression, the flux of the FS EIC emission would be of the
form \citep{Murase+2011EIC}:
\begin{equation}
 F_\varepsilon^{FSEIC} \propto \left\{
\begin{array}{ll}
 \varepsilon^{1-\alpha}
& \textrm{if}\, \varepsilon \leq \varepsilon_{EIC,c}\\
 \varepsilon^{-1/2}
& \textrm{if}\, \varepsilon_{EIC,c} < \varepsilon \leq \varepsilon_{EIC,m}\\
 \varepsilon^{-p/2}
& \textrm{if}\, \varepsilon_{EIC,m} < \varepsilon
\label{eq:fs-eic-flux}
\end{array}
\right.
\end{equation}
The KN effects however would introduce breaks in the high-energy part of the
spectrum. Details of the derivation of the KN frequencies are in Appendix
\ref{KN}.  For a large part of the parameter space, $\ve_{EIC,m}$ will be in
the KN regime, and the $\ve F_\ve$ peak will be at the KN break {frequency
$\ve^{EIC}_{KN}\approx \eta \gamma_c m_e c^2/(1+z)=15~ (1-\zeta_r)^{-1/3}
L_{t,53}^{-1/3} t_{1.3}^{-1/3} \epsilon_{B,-1}^{-1} n_0^{-2/3} \eta_{600}^{2/3}
(1+Y)^{-1} ((1+z)/2)^{-1} \GeV$.}   A rough estimate of the peak of the EIC
emission at the peak is
\begin{eqnarray}
&(\varepsilon F_{\varepsilon})_{EIC}^{\rm peak} \sim \varepsilon_{br}
N_{\varepsilon,p} \tau_{FS} \ve^{EIC}_{KN} =\\
7.2\times 10^{-10} &  L_{t,53} \zeta_r^{3/2} (1-\zeta_r)^{-1/2}
\epsilon_{B,0}^{-1/2} \epsilon_{B,-1}^{-1}
 \Gamma_r^{-3} (1+Y)^{-1} D_{L,28.2}^{-2}
{\fflunit},
\label{eq:fs-eic-pk}
\end{eqnarray}
where $N_{\varepsilon,p}$ is the photon number spectrum of the prompt emission,
evaluated here for $\alpha=1$, $\beta=2.4$ and for $\ve\approx\ve_{br}$.  To
account for the anisotropy of the emitted radiation in the forward shock frame,
we multiply our flux by a factor of $0.5$ \citep{fanpiran06b}.

\subsection{EIC scattering of photospheric non-thermal photons on external reverse
shock electrons (RS-EIC)}
\label{sec:rs-eic}

If the reverse shock develops, there will be an external inverse Compton
component from the reverse shock electrons scattering prompt photons as well.
This differs from the forward shock EIC because of the larger optical depth and
the lower energy of the electrons in the RS, and because the RS will be in the
slow cooling phase.
The characteristic frequencies are $\varepsilon_m^{RSEIC} \approx 2
\gamma_{RS,m}^2 \varepsilon_{br} = 17 ~ \epsilon_{e,-2} g_{p,2.4} r_{0,7}
\epsilon_{B,0}^{1/2}   \Gamma_{r}^3 ((1+z)/2)^{-1} \MeV$ and $
\varepsilon_c^{RSEIC} \approx 2 \gamma_{RS,c}^2\varepsilon_{br} = 6.0  ~
L_{t,53}^{-2/3} \zeta_r^{-1/2} (1-\zeta_r)^{-1/6}   t_{1.3}^{-2/3}
\epsilon_{B,0}^{1/2} \epsilon_{B,-1}^{-2} n_0^{-4/3} \eta_{600}^{-2/3}
(1+Y)^{-2} r_{0,7}^{-1/2}\Gamma_r^3 ((1+z)/2)^{-1} \GeV$.
The RS electrons, being in the slow cooling phase, will upscatter the prompt
emission into a spectrum of the following shape:
\begin{equation}
 F_\varepsilon^{RSEIC} \propto \left\{
\begin{array}{ll}
 \varepsilon^{1-\alpha}
& \textrm{if}\, \varepsilon \leq \varepsilon^{RSEIC}_{m}\\
 \varepsilon^{(1-p)/2}
& \textrm{if}\, \varepsilon^{RSEIC}_{m} < \varepsilon \leq \varepsilon^{RSEIC}_{c}\\
 \varepsilon^{-p/2}
& \textrm{if}\, \varepsilon^{RSEIC}_{c} < \varepsilon.
\end{array}
\right.
\label{eq:rs-eic-flux}
\end{equation}

The peak of the emission will be at the Klein Nishina cutoff frequency for this
component, which is $\varepsilon_{KN}^{RSEIC}\simeq \eta \gamma_m^{RS} m_e c^2
/(1+z)\approx 0.8 ~\eta_{600} \epsilon_{e,-2} g_{p,2.4} ((1+z)/2)^{-1} \gev$.
This introduces a spectral break and the photon index above this energy will be
$\sim -(\alpha-p-2)$.

The peak flux of the emission, considering a weakening by 0.5 due to anisotropy, is
\begin{equation} \varepsilon  F_{\varepsilon,RSEIC}^{{\rm peak}}  \approx
\varepsilon_{br} N_{\varepsilon,p} \tau_{RS} \varepsilon_{KN}^{RSEIC} =
2.3\times 10^{-8}~ L_{t,53}^{4/3}  \zeta_r^{3/2} (1-\zeta_r)^{-1/6}
t_{1.3}^{1/3}  \eta_{600}^{4/3} n_0^{2/3}  \Gamma_r^3 r_0^{-1/3}
\epsilon_{B,0}^{-1/2} \epsilon_{e,-2} g_{p,2.4}  D_{L,28.2}^{-2}{\fflunit}.
\label{eq:rs-eic-pk} \end{equation}
{Note that $\varepsilon^{RSEIC,m}_{KN}$ falls  in the low-energy part of
the {\it Fermi} \lat range.  Thus, if the RS-EIC were dominant, the photon index
expected would be $-(\alpha-p-2)\approx3.4$. In bursts with extra high-energy
components, such soft photon indices (or even softer) are indeed observed.  For
other model parameters, we found that while the RS-EIC	is  the dominating
component at its peak, other inverse Compton radiation components such as
FS-EIC of FS-SSC can make the spectrum harder by contributing at a few
$\times10 \gev$ s.  }

\section{Numerical Results and Model Parameter Variations}
\label{sec:results}

We calculated a number of model spectra based on the considerations of the
previous sections. The initial set of nominal parameters used is $L_{t}=
10^{53}~ {\rm erg/s}$, $t=20$ s, $\zeta_{r}=0.5$, $\zeta_{k}=0.5$, $n=1~{\rm
cm}^{-3}$, $\eta=600$, $\epsilon_{B,pr}=1$,
$\epsilon_{B,FS}=\epsilon_{B,RS}=0.1$ , $\epsilon_{e,FS}=\epsilon_{e,RS}=0.01$,
$r_0=10^7~{\rm cm}$, $ z=1$ , $D_L\approx2\times10^{28}~ {\rm cm}.  $

Here $\zeta_r=0.5$ is the nominal fraction of the initial $L_t$ assumed to be
radiated at the photosphere in non-thermal and thermal	radiation,
$\zeta_{k}=0.5$ is the energy radiated by the external shock at the
deceleration radius Also, for simplicity, when a reverse shock forms, we assume
the same values for $\eps_B$ and $\eps_e$ in both the forward and in the
reverse shocks.  We have explored also the effects of departures from these
various nominal parameters.

\noindent
{\it Models with Negligible Pair Formation}.- A first set of models was
calculated assuming that the first scenario of \S \ref{sec:nonth}, where the
photospheric spectrum of eq. (\ref{eq:Band}) cuts off around $m_e c^2$ in the
jet frame, or around $\sim 50$ MeV in the observer frame, with pair formation
being negligible in both the photosphere and in the external shock.  The
resulting generic observer-frame spectrum consists of a Band function spectral
component  peaking at sub-MeV energies and extending up to $\sim 50$ MeV, with
a second component at GeV energies due to a combination of EIC and SSC by
external forward and reverse shock electrons of photospheric photons and their
own synchrotron photons. This second GeV component has typically a total
fluence which is $\siml 0.1$ that of the MeV Band function, and for some
parameter ranges it stands out from the first component, while for others it
merges more or less smoothly with the high energy branch of the first
component. Whether it stands out or not depends on the external density, the
external shock parameters, and on whether a reverse shock forms or not.

\begin{figure}[htbp]
\begin{center}
\includegraphics[width=1.0\columnwidth]{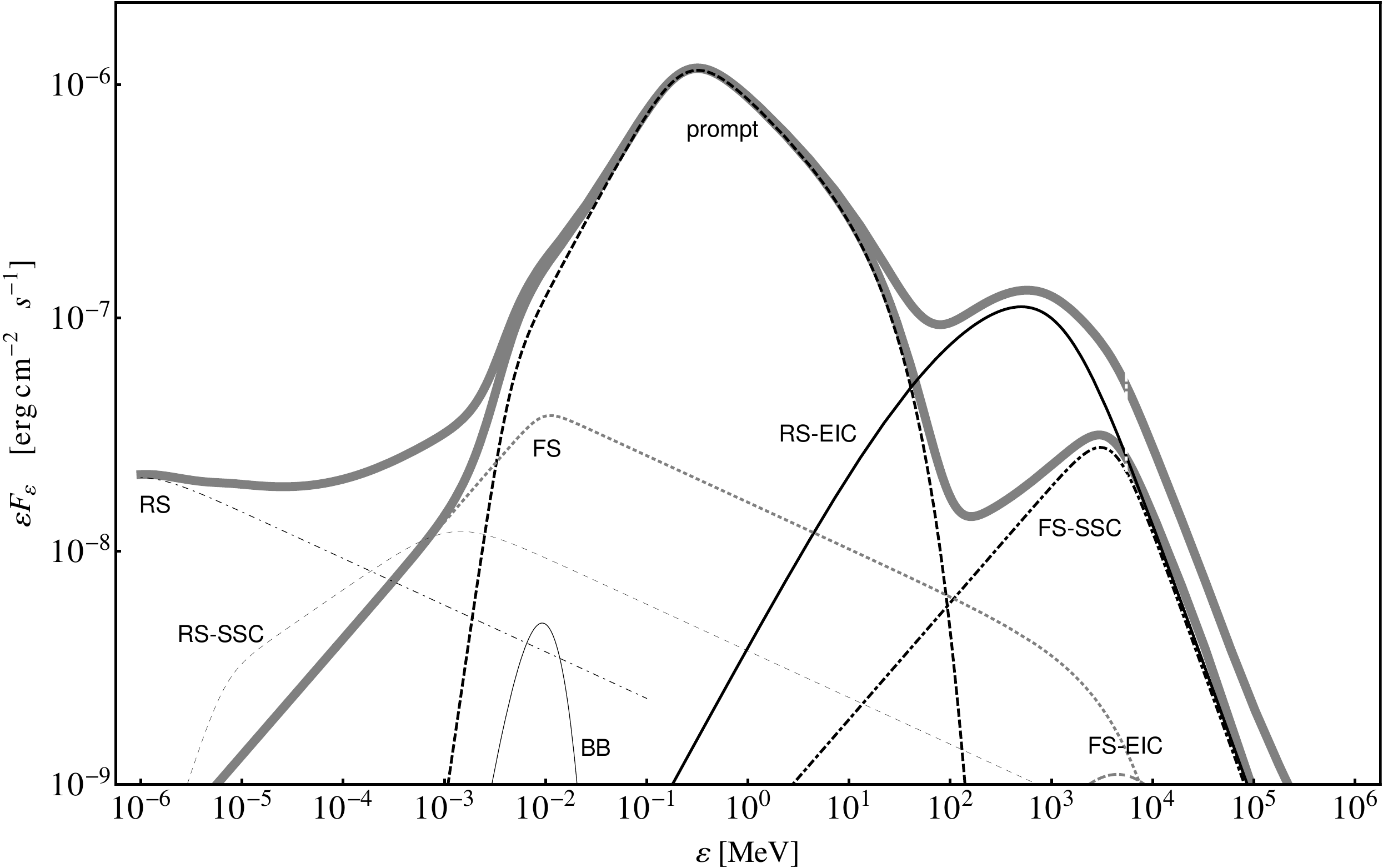}
\caption{A model without pair formation, $L_{t}=5\times 10^{52}~ {\rm erg/s},
t=20~{\rm s}, \zeta_{r}=0.6, \zeta_{k}=0.4, n=100~{\rm cm}^{-3},~ \eta=400,
\epsilon_{B,pr}=0.9, \epsilon_{B,FS}=1\times10^{-2},~
\epsilon_{e,FS}=2\times10^{-2}, r_0=10^7~{\rm cm}, z=1, \beta=2.4, p=2.4$.  The
black dashed line is the prompt synchrotron emission, black thin continuous
line is the prompt thermal component (marked BB), the thick black line is the
RS-EIC, the gray, thick, dotted line is the forward shock synchrotron part
(FS), the gray, dashed line is the forward shock external inverse Compton, the
black dash-dotted is the FS-SSC component, the gray, dashed line is the RS-SSC
and the gray, dash-dotted is the RS synchrotron component.  The thick gray
continuous line is the sum of the components (the upper one is with and the
lower one without the RS contributions).  } \label{fig:model-1} \end{center}
\end{figure}

In Figure \ref{fig:model-1} we show one of the cases where the second, GeV
component, stands out from the first Band component. The parameters of the
external forward shock are $\epsilon_{B}=0.01$ and $\epsilon_{e}=0.02$, and the
dissipative photosphere produces a Band function peaking at $\lesssim$ 1 MeV in
the {\it GBM} range.

In the {\it LAT} range, if a reverse shock is present, we can see a clear
``bump" in the spectrum around $1 \GeV$ (the upper thick curve).  The prompt
emission and the bump up to $\sim 3 \GeV$ in Figure \ref{fig:model-1} is
strikingly similar to the spectrum of GRB 090926A \citep[][their figure
5.]{Ackermann+11-090926}, which showed an extra power law with a cutoff besides
the Band component. Here the extra power law would be the rising part of the
RS-EIC (the model photon index $(-(-p-1)/2\simeq1.7$ {coincides precisely with
the measured one for $p=2.4$}) and the cutoff is the part after the RSEIC peak.
The ratio between the \gbm and \lat fluences is $\sim 10^{-1}$, an average
ratio in the observed bursts \citep{Omodei+11fermigrb, Pelassa+11-latgrbrev}.
At optical wavelengths the reverse shock produces a high flux ($\sim
3\times10^{-8} \fflunit$, or $m_R\sim 7$) which could be responsible for the
very bright optical flashes observed in some bursts.

However, if the reverse shock is weak or missing, in the same figure
\ref{fig:model-1} the upper thick bump at $\sim 1 \GeV$ is absent, and is
replaced by the lower thick line.  One has the same {\it GBM} prompt emission
and a hard component emerging at $\lesssim10 \gev$.  This component would also
appear as an extra power-law component, but in this case, it is weaker by one
order of magnitude than the prompt emission. For the low photon number detected
in this energy range, such a component might have low significance in a fit.
Also the optical flux is much lower, $3\times 10^{-10} \fflunit$, or $m_R \sim
11.8$. We note that the combination of parameters used for obtaining the single
component and two component spectra in this figure are not unique.

\begin{figure}[htbp]
\begin{center}
\includegraphics[width=1.0\columnwidth]{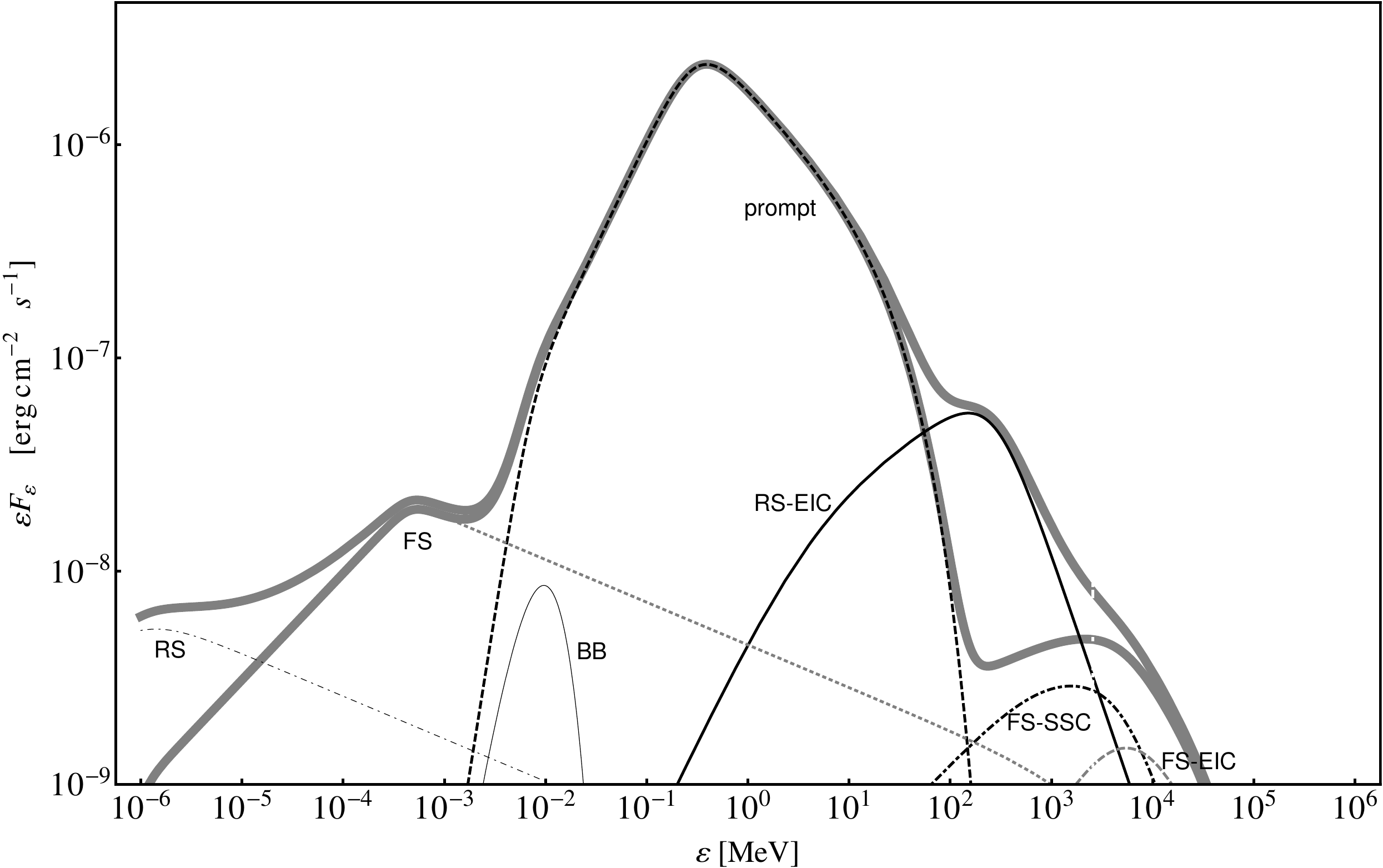}
\caption{A different model, also without pair formation, and parameters $L_{t}=
10^{53}~ {\rm erg/s}, t=20~ {\rm s}, \zeta_{r}=0.5, \zeta_{k}=0.5, n=30~
{\cm}^{-3}, \eta=400, \epsilon_{B,pr}=1, \epsilon_{B,FS}=\epsilon_{B,RS}
=2\times10^{-2}, \epsilon_{e,FS}=\epsilon_{e,RS}=5\times10^{-3}, r_0=10^7~{\rm
cm}, z=1, \beta=2.5, p=2.4$.  The black dashed line is the prompt synchrotron
emission, black thin continuous line is the prompt thermal component (marked
BB), the thick black line is the RS-EIC, the gray, thick, dotted line is the
forward shock synchrotron part (FS), the gray, dashed line is the forward shock
external inverse Compton, the  black dash-dotted is the FS-SSC component and
the gray, dash-dotted is the RS synchrotron component.	The thick gray
continuous line is the sum of the components (the upper one is with and the
lower one without the RS contributions).  } \label{fig:model-2} \end{center}
\end{figure}

In Figure \ref{fig:model-2} we show one of the cases where the overall spectrum
appears as a single Band component, due to the second GeV component approximating
an extension of the first, MeV Band component.
The parameters of the external shock are in this case $\epsilon_{B}=0.02,~
\epsilon_{e}=0.005$ for both the forward and reverse shock, with $\eta=400$ and
$n=30~{\cm}^{-3}$, other parameters being nominal. It is seen that the
combination of the external reverse shock EIC and the forward shock SSC plus
EIC provides a roughly continuous slope connecting the MeV Band spectrum to the
high energy branch (the upper thick curve from $\sim 0.1 \GeV$ onwards).  The
reverse shock EIC hump fills in the through left by the forward shock SSC.  The
wiggles from the two humps at 100 MeV and 10 GeV would be smoothed out in
typical spectral fits with low photon number statistics in this range
encountered in \fermi\lat bursts. The matching of the flux levels between the
photospheric spectrum high energy end and the external RS peak is controlled
largely by the external density $n$, whose value here is in the usual range.

Assuming that a reverse shock is absent or negligibly weak (i.e. ignoring the
upper thick gray continuous line in fig. \ref{fig:model-2} peaking at $\gtrsim 100
\MeV$), one sees that now a second hard component appears above 100 MeV and
peaks at 3 GeV, being in this case mainly due to the forward shock SSC, with a
contribution from the EIC of photospheric photons.  Other components, while
present, are here sub-dominant.  The secondary peak (also in the absence of the
RS-EIC component in fig. \ref{fig:model-2}), though reminiscent of the extra
power-law component observed in GRB 090902B or 090926A, is $\sim2$ orders of
magnitude fainter than the main peak, while in the observed cases this ratio,
when reported, is at most $\sim1.5$. The optical flux predictions {are
$m_R\sim 8.3$ and $m_R\sim 10.5$ in the presence or absence of the reverse
shock.} Again, this is not the sole combination of parameters which produces an
approximate single Band function extending to GeV energies.

In the above models where pair formation is not expected, the qualitative
effects of increasing the terminal Lorentz factor $\eta$ consists in a
strengthening of the RS-EIC component. Even though at low $\eta$ we expect a
lower cutoff for the prompt emission and a more prominent inverse Compton
component, this is not the case.  Below $\eta \simeq 300$ the main high-energy
component is the sum of the FS-SSC, the FS-EIC and the prompt SSC at $\sim 10
\GeV$ which form a distinct peak.
Increasing the density $n$ makes the RS-EIC component to be more detectable
from $n\simeq10 \cm^3$ up to $n\simeq500 \cm^3$. Outside these parameters the
FS-SSC and the FS-EIC are the same magnitude or dominating the RS-EIC component
and a break in the spectrum is more visible.
The shape of the high energy part of the spectrum has only a weak dependence on
the value of $\epsilon_B$. By decreasing the prompt magnetic parameter, the
peak energy becomes lower and the external components could become more
prominent.

\noindent
{\it Models with Pair Formation}.-
A second set of models was calculated assuming that the Band spectrum from the
photosphere extends to sufficiently high energies that pair formation from
$\gamma\gamma$ interactions is expected (see \S \ref{sec:phot}). In these cases
the primary Band component extends up to an energy given by
eq.(\ref{eq:ephgg}), and there is a secondary synchrotron component from the
$\gamma\gamma$ pairs, both the primary and secondary photons leading to
separate  SSC and EIC components from upscattering in the photosphere and in
the forward and reverse external shock, some components being more important
than others.
\begin{figure}[htbp]
\begin{center}
\includegraphics[width=1.0\columnwidth]{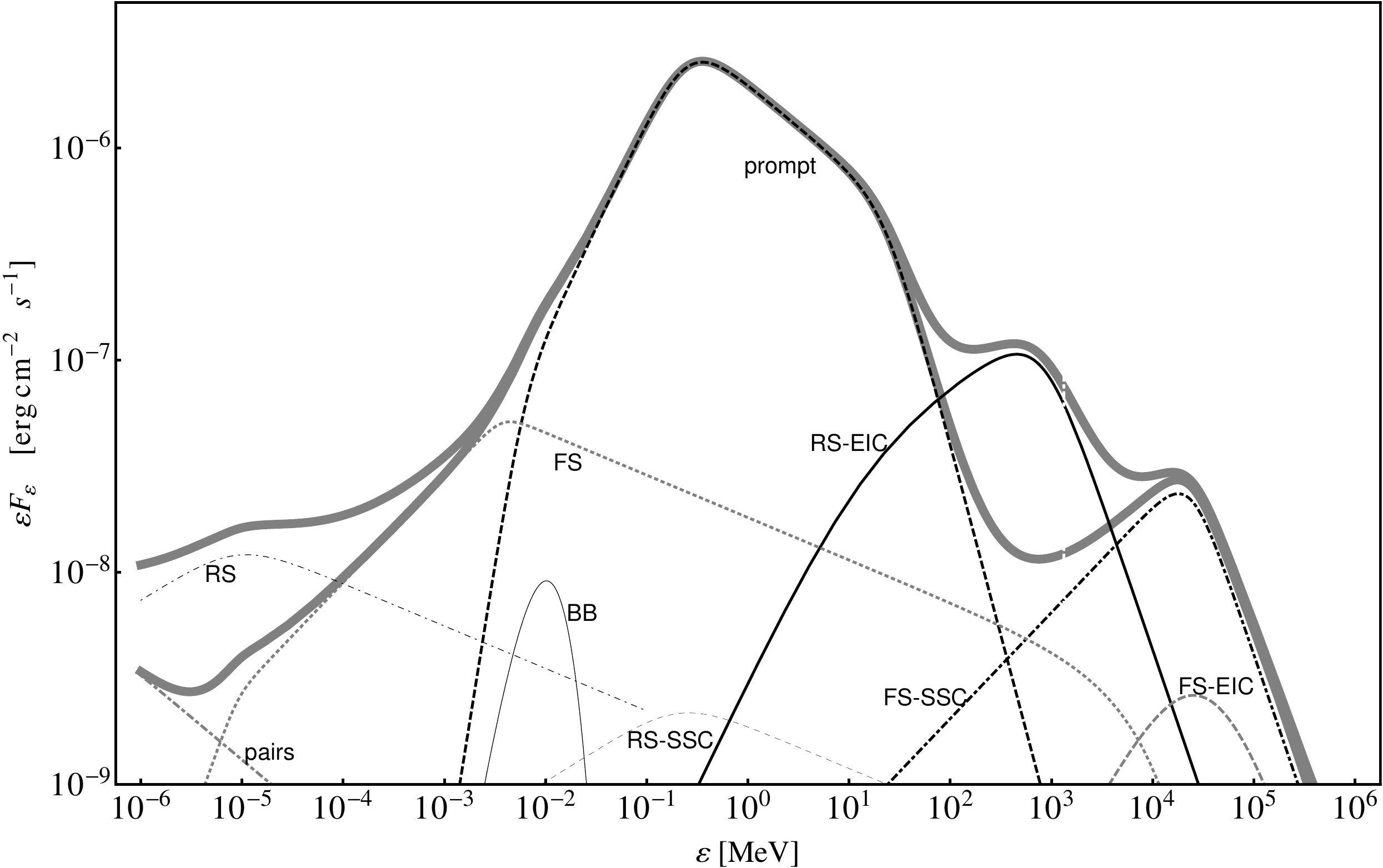}
\caption{Model with pair formation with $L_{t}= 10^{53}~{\rm erg/s}, t=20~ {\rm
s}, \zeta_{r}=0.6, \zeta_{k}=0.4, n=10~ {\rm cm}^{-3}, \eta=600,
\epsilon_{B,pr}=1, \epsilon_{B,FS}=\epsilon_{B,RS}=1\times10^{-2},
\epsilon_{e,FS}= \epsilon_{e,RS}=1\times10^{-2}, r_0=10^7~{\rm cm}, z=1,
\beta=2.4, p=2.4$.  The black dashed line is the prompt synchrotron emission,
black thin continuous line is the prompt thermal component (marked BB), the
thick black line is the RS-EIC, the gray, thick, dotted line is the forward
shock synchrotron part (FS), the gray, dashed line is the forward shock
external inverse Compton, the  black dash-dotted is the FS-SSC component, the
gray, dash-dotted is the RS synchrotron component and the thick dash-dotted is
the pair synchrotron contribution.  The thick gray continuous line is the sum
of the components (the upper one is with and the lower one without the RS
contributions).  } \label{fig:model-3} \end{center} \end{figure}
Figure \ref{fig:model-3} presents one of the cases involving pair production.
The overall behavior is not too different from that of the models with no pair
production: the RS-EIC component is approximately smoothly joined to the prompt
at $\sim 100 \MeV$. The RS-EIC in turn joins smoothly to  the other higher
energy components at $\sim 10\GeV$. If a RS does not develop, the only missing
component {at high energies} will be the RS-EIC (thick black line) and again we
get a bump in the spectra at $\sim 10 \GeV$ energies. The optical flux is
{only} about $1.3$ magnitudes fainter in the absence than in the presence of
the reverse shock, {$m_R\sim9.3$ and $\sim8$ for the two cases.}

\begin{figure}[htbp]
\begin{center}
\includegraphics[width=1.0\columnwidth]{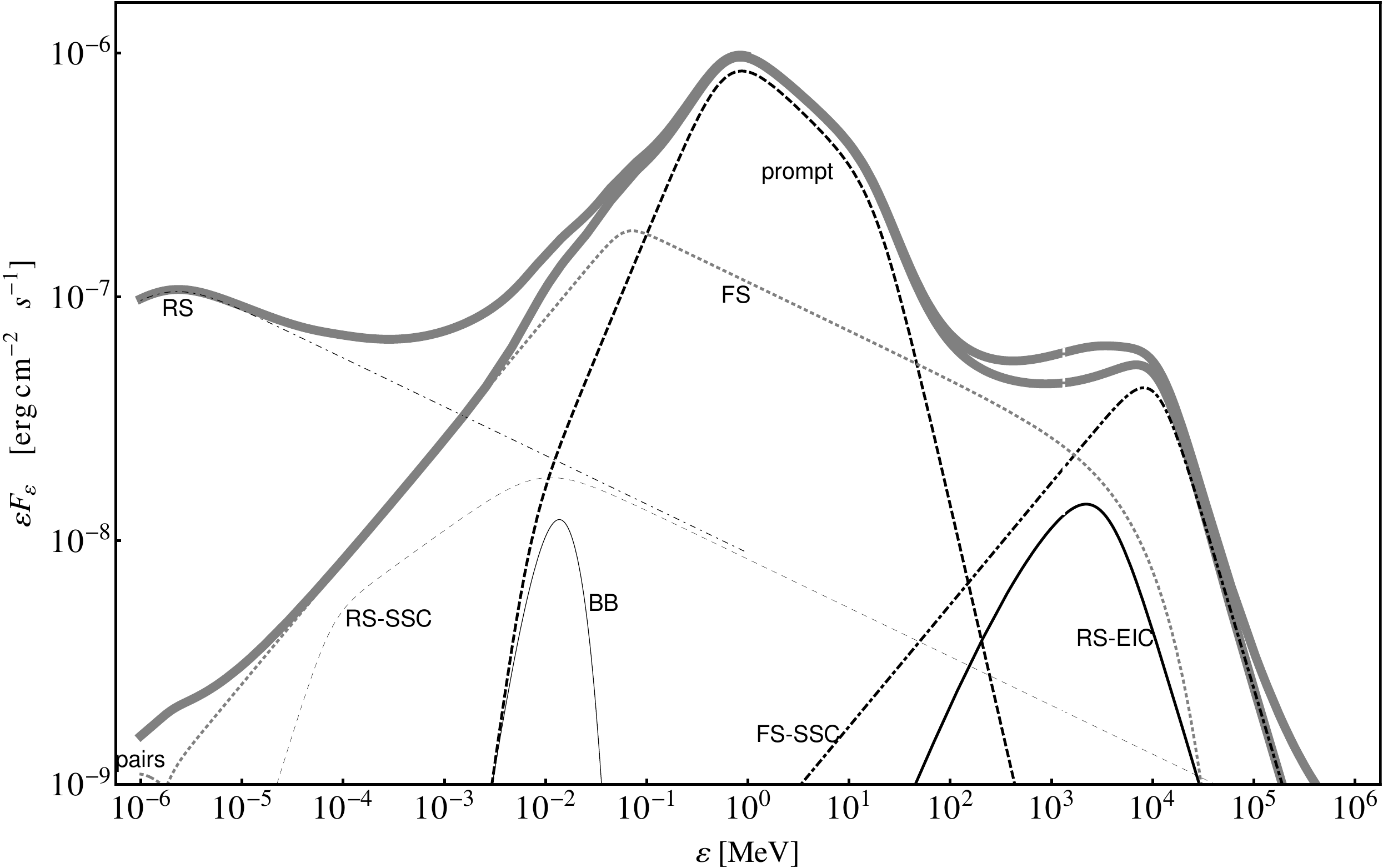}
\caption{Another model with pair formation, and parameters $L_{t}= 10^{53}~{\rm
erg/s}, t=20~ {\rm s}, \zeta_{r}=0.2, \zeta_{k}=0.8, n=1~ {\rm cm}^{-3},
\eta=600, \epsilon_{B,pr}=1, \epsilon_{B,FS}=\epsilon_{B,RS}=0.1,
\epsilon_{e,FS}=\epsilon_{e,RS}=1\times10^{-2}, r_0=10^7~{\rm cm}, z=1,
\beta=2.4, p=2.4$.   The black dashed line is the prompt synchrotron emission,
black thin continuous line is the prompt thermal component (marked BB), the
thick black line is the RS-EIC, the gray, thick, dotted line is the forward
shock synchrotron part (FS), the gray, dashed line is the forward shock
external inverse Compton, the  black dash-dotted is the FS-SSC component, the
gray, dash-dotted is the RS synchrotron component and the thick dash-dotted is
the pair synchrotron contribution.  The thick gray continuous line is the sum
of the components (the upper one is with and the lower one without the RS
contributions).  } \label{fig:model-4} \end{center} \end{figure}

In Figure \ref{fig:model-4} we show another case where pair formation occurs,
for a different choice of parameters. In this case the second component bump
would be present irrespective of whether the reverse shock is present or
absent, since both shocks result in a similar bump. {The FS synchrotron and the
FS-EIC	contribute significant flux to the bump. } In this case, the optical
flux with the reverse shock is larger, $m_R\sim 5.5$, and much fainter in the
absence of the reverse shock, $m_R\sim 10$.

\noindent
{\it Low Energy Power Law Extensions}.-
A notable feature of Figs. \ref{fig:model-1} through \ref{fig:model-4} is that
the forward shock synchrotron radiation extends into the optical range.  In
\citet{Asano+10optex}, such a power law extending into the optical was obtained
for a hadronic cascade GRB model. Here it arises in a purely leptonic mode.	For
reasonable parameters this component falls between the cooling and the
characteristic synchrotron frequencies.  One sees that, for bright \fermi \lat
bursts, in the absence of pair formation the forward shock synchrotron can
produce a prompt optical flash of $m_R \gtrsim 11-12$, even in the absence of a
reverse shock. This visual flux scales with the prompt photon luminosity $L_r$.
A conversion of the flux units to  R-magnitudes is $10^{-8}$ erg cm$^{-2}$
s$^{-1}$~$\to ~m_R\sim 8.0$, and $10^{-10}$ erg cm$^{-2}$ s$^{-1}$~$\to
~m_R\sim 13.0$, with the usual five magnitudes interval per each factor 100
increase in flux.  In the cases where pair formation occurs, the pairs
contribute an additional synchrotron component, which in the optical range
predicts an even brighter prompt flash. This is seen in Fig. \ref{fig:model-3},
with the source at $z=1$. In both types of models (with or without pair
formation), in the cases where a reverse shock is present the optical flashes
can be significantly brighter, in the range $m_R\sim 6-8$ at the same redshift.

\section{Discussion}
\label{sec:disc}

We have addressed the high-energy spectral properties of the bursts observed
with Fermi \lat,  using a magnetically dominated outflow model where the prompt
MeV emission arises in the photosphere, and high energy components arise from
inverse Compton scattering by both photospheric and external shock electrons.
We have investigated circumstances under which a single Band function appears
to extend to the highest energies detected by the \fermi\lat, and where a second
high energy component shows itself above the MeV range Band spectrum. We have
also addressed, in the same context, the production of prompt optical flashes.

We find that, qualitatively, these types of models are able to explain the observed
diversity of \gbm and \lat prompt emission spectra, without invoking an
internal shock emission at intermediate radii. The presence or relative strength
of a reverse shock plays a role in determining the spectral shape and the basic
dichotomy in the latter. As discussed by, e.g. \citet{Narayan+11magn,
Mimica+11magshock,Giannios+08rsmag}, for modest magnetization $\sigma$ of
the outflow a reverse shock may be expected; and while for initially
magnetically dominated outflows $\sigma$ (or $\epsilon_B$) is still large at
the photosphere, at larger radii it can be expected to drop sufficiently low to
make reverse shocks possible, although uncertainties remain concerning the
threshold magnetization and the reverse shock strength. Thus, we assumed that
within the normal range of parameters, reverse shocks may be important in some
bursts and not in others.  When a reverse shock is absent, the forward shock
results in a distinct high energy (GeV) spectral component, whose fluence is
{1-2} orders of magnitude below that of the Band component. In these cases, the
effective high energy slope $\beta$ of the Band component appears rather soft,
in agreement with data discussed in, e.g. \citet{Zhang+11-latgrb}. On the
other hand, when a reverse shock is present, its inverse Compton radiation can
result in a larger fluence extra high energy component, 0.5-1 orders of
magnitude below the Band fluence; or it can result in a smooth continuation of
the Band component, which can mimic a single Band high energy component of
relatively hard effective slope $\beta$. The relatively small number of photons
observed in this band could result in fits where the slight wiggles in the
theoretical spectra are largely ironed out.

The fact that the \lat emission is contributed by the external shock implies
that it will appear with an intrinsic delay of order $t_{del}\sim r_{dec}
(1+z)/c\eta^2 \approx4.4~ L_{t,53}^{1/3} (1-\zeta_r)^{1/3} t_{1.3}^{1/3}
n_0^{-1/3} \eta_{600}^{-8/3} (1+z)/2~ {\rm s}$, of order a few seconds relative
to the photospheric MeV component. Such a delay was indicated also in some
early work on \lat spectra \citep[e.g.,][]{Ghisellini+09-gevrad,
Kumar+09LATexternal}, where, however, the \lat emission was attributed to the
forward shock synchrotron radiation.  In our case, it is the inverse Compton
components of the forward or reverse shock which dominate the \lat emission.

The implications of our model for constraints on the bulk Lorentz factor are
much less stringent than in one-zone models where the \gbm and \lat emission
are assumed to arise in the same region. E.g., in \citet{Abdo+09-090902B,
Ackermann+11-090926, Fermi+09-080916-sci}, such analyses indicated Lorentz
factors $\gamma \sim 800-1000$ or higher.  However \citep{Zou+11Lorentz,
Peer11-fermigrb, Zhao+11lorentz} in generic two-zone models the Lorentz factor
need not be so large. Specifically, in our model, which is a two-zone model in
which the high energy photons arise in the external shock, the compactness
parameter in the latter is low, and the spectra can be reproduced with terminal
Lorentz factors $\Gamma\sim \eta\sim 300-600$.

As discussed in \S \ref{sec:nonth}, the magnetized dissipative photosphere can
produce a Band-like non-thermal spectrum resembling the observations.  In our
magnetized models, in the absence of pair formation this component cuts off
above $\sim 50 \MeV$, or in the presence of pair formation, it steepens by one
power law index above $\sim 100\MeV$.  Recently \citet{Kocevski+12LATpaucity}
analysed the paucity of GRBs measured by LAT. They claim that nearly half of
the bursts detected by GBM which were in the LAT field of view required a break
under $0.1 \GeV$ to explain the nondetection by LAT.  In the framework of this
model  these results mean that there is indeed a cutoff at $\sim 50-100 \MeV$.
This is either	due to a softening because of  pair creation or a cutoff according
to the magnetic acceleration mechanism.

Recently \citet{Yonetoku+11pol} reported a polarization measurement in the
prompt emission from GRB 100826A.  This could be an indication of a magnetically
dominated photosphere \citep{Waxman03gammapol,Nakar+03gammapol}, although
polarization might also be expected from processes not requiring strong magnetic
fields \citep[e.g.,][]{Lazzati+04dragpol}. For a magnetic jet, the transverse
field components will dominate in the emission region, and while for an observer
line of sight along the jet axis the polarization could average itself out, for
the larger probability off-axis viewing directions a net polarization could be
expected.

The magnetic photosphere models also predict a weaker thermal component peaking
at a few keV (\S \ref{sec:th}), plotted in Figs. \ref{fig:model-1} {
through} \ref{fig:model-4}.  Such a thermal component has been reported in
\cite{Guiriec+10-100724} for GRB 100724B. In our models such a component
appears at approximately the right energies, its fluence generally being low
compared to the nonthermal components.	However, only for a relatively small
range of parameters would it appear possible to detect it. One problem is that
it can be conflated with the contribution of the forward shock synchrotron,
e.g. as the bump around $\lesssim10$ keV in Fig.  \ref{fig:model-3}.

Another interesting component is the optical band extension of the external
shock synchrotron spectra (see Figs. \ref{fig:model-1} through
\ref{fig:model-4}).  It is seen that just the external forward shock by itself
already can produce optical flashes of $m_R \gtrsim 12$ (Figs.
\ref{fig:model-1}, \ref{fig:model-2}), while if pair formation occurs in the
photosphere, the cooled pairs there can lead to flashes of $m_R \sim 9.3$
(Figs. \ref{fig:model-3}, \ref{fig:model-4}) or even brighter for suitable sets
of parameters.  On the other hand, when a reverse shock is present, its
synchrotron component naturally produces a bright prompt optical flash, as
known for quite a while \citep[e.g.,][]{1993ApJ...418L..59M,Meszaros+97ag}.
Here, in addition, we have considered also the IC components of the reverse
shock, and the effect of a photospheric EIC component as well.	For reasonable
parameters, the flux can be close to few $\times10^{-7} \fflunit$ in the
optical band, which translates to $m_R$ brighter than $7$ at $z=1$.
Generically, the relative scarcity of observed optical flashes may be
attributed to the fact that their brightness scales roughly the same way as the
prompt GeV luminosity ($(\ve F_\ve)_{\rm peak}^{RS}\propto L_t^{p/3} \propto
L_t^{0.8}, (\ve F_\ve)_{\rm peak}^{\rm prompt}\propto L_t $ and $(\ve
F_\ve)_{\rm peak}^{RS-EIC}\propto L_t^{4/3} $ ), as well as to the fact that
reverse shocks may be rare in magnetically dominated outflows.

For the naked eye GRB 080319, the roughly similar behavior of the optical and
$\gamma$-ray light curves can be used to argue for a common origin of both
\citep{Swift+08-080319}. However, the optical light curve is not sampled
as well as the $\gamma$-ray light curves, and at least in the prompt phase
 shows temporal structures (peaks) comparable in duration  to the
deceleration time   which could be compatible with a reverse shock origin,
although the RS origin was disfavoured by \citet{Swift+08-080319} .
On the other hand, a fast variability of the prompt optical flash might be
suggestive of an origin in the same region as the prompt MeV emission, which
might be attributed to the photospheric cooled pair synchrotron component.
However, only very rare parameter combinations could push the optical flux of
the pair synchrotron component up to $m_R\sim 5$ as in the naked eye burst
\citep{Swift+08-080319}. E.g. in our model a combination of $\eta\sim 1000$,
$\beta\approx2.1$ and $\zeta_r\approx0.99$ would approach such brightnesses.

%ref valasz (4)
 Concerning the FS-EIC and RS-EIC components, strictly speaking for these
it is not necessary to invoke the magnetically dominated jet model. Indeed,
similar results can be obtained by a baryonic model as well \citet{Beloborodov05gev,
Beloborodov10phot}. In magnetic and baryonic cases the delay between the MeV and
GeV components will be a few seconds, comparable to the values measured by \fermi.

We note that this model may be applicable both to long and short bursts, since a
magnetized photosphere and an external shock would be expected in both cases.
The relatively shorter GeV-MeV delays in the short burst cases could be
understood in terms of a closer-in deceleration or a larger Lorentz factor, the
latter being suggested also by their harder MeV spectra.

Finally, we point out that the predicted inverse Compton components in these
models extend into the TeV range, for a range of parameters. The photon with
the highest energy detected by \lat from a GRB had an energy of $\sim 33$ GeV
\citep{Abdo+09-090902B}.  However, the spectral features above this energy are
in the range of ground-based Cherenkov telescopes, including also HAWC and the
future CTA, providing potentially interesting targets for such detectors.

\acknowledgements{We acknowledge NASA NNX09AL40G, NSF PHY-0757155 and OTKA
grant K077795 for partial support, and thank Bin-Bin Zhang, Shan Gao, Kenji
Toma and the referee for useful comments.

\eject
\appendix
\section{Other Radiation Components}
\label{sec:app-proton}

{ Other electron inverse Compton components, besides those already discussed,
may be present if a reverse shock develops. In this case one would expect
scattering of reverse shock photons on the forward shock electrons, and FS
photons on the RS electrons. Both of these components have a flux density
\citep{2011ApJ...733...22H} $ \tau_{FS} F_{\varepsilon_{\max}}^{RS}\approx
\tau_{RS} F_{\varepsilon_{\max}}^{FS}\approx  10^{-6}\ {\rm Jy} $ peaking at
$\varepsilon \approx 0.1 \keV $ with a flux of $(\ve F_\ve)_{\rm peak} \approx
2\times 10^{-13} \fflunit$, which  makes them negligible  compared to other
(e.g. FS or RS ) components.}

The external inverse Compton radiation of the pair synchrotron photons on the
FS and the RS will also give a flux of the order $5\times 10^{-10} \fflunit$ at
$80 \MeV$ and $0.4 \MeV$ respectively.

The prompt thermal (blackbody) component will also be upscattered at the FS as
well at the RS	\citep{2008ApJ...689..351A}.  Both components consist of a set
of smoothly joined power-laws and a cutoff at high energies.	The BB-FS-EIC
has a peak of  $9\times10^{-12} \fflunit$ at $\sim 30\gev$ while the BB-RS-EIC
has a peak of  $6\times10^{-12} \fflunit$ at $\sim 4 \gev$.

The synchrotron self Compton component of the reverse shock peaks (in $\ve
F_\ve$) at $\varepsilon_{c,RS}^{SSC} \approx 2\gamma^2_{c,RS}\varepsilon_{c,RS}
\approx35 \keV$. The Compton Y parameter, which gives the luminosity ratio of
the SSC to the synchrotron component is $Y_{RS}^{SSC} = (-1 + \sqrt{1 + 4
\eta_{RS}^{SSC} \epsilon_e/ \epsilon_B}) / 2 \approx 0.031$
\citep{2001ApJ...548..787S}, where $\eta_{RS}^{SSC}=\min
((\gamma_c^{RS}/\gamma_m^{RS})^{2-p}, 1)$. In this slow cooling regime,
$\eta_{RS}^{SSC}<1$ is valid.  The amplitude of the RS-SSC is obtained from
$(\ve F_{\varepsilon})_{SSC}^{\rm peak} = Y_{RS}^{SSC} \varepsilon^{RS}_c
F_{\varepsilon_c}^{RS}\approx 3.5\times 10^{-10}\fflunit$.  While this
component can have fluxes of the order of $10^{-8}\fflunit$, it is dominated by
other components.

The prompt emission will generate an SSC component. The peak energy of
this component $2 \gamma_{e,ph}^2 \ve_{br}\approx 0.2~ {\rm TeV}$ falls in the
deep KN regime and its $\ve F_\ve$ peak will be less than $10^{-9} \fflunit$.

\section{Klein-Nishina break for SSC and EIC components}\label{KN}
\label{sec:app-kn}

Klein-Nishina effects are potentially important when assessing high-energy,
inverse Compton components. In our study we investigate the KN break for SSC
and EIC components of both the FS and RS with the prompt emission.  We
calculate the KN break energy based on \citet{Guetta+03plerion}.  The KN break
occurs at the solution of the $\varepsilon=\gamma_{max}(\varepsilon) m_e c^2$,
where $\gamma_{max}$ is a function of $\varepsilon$ \citep[for the detailed
expression of $\gamma_{max}(\varepsilon)$ see][]{Guetta+03plerion}.  Depending
on the radiative regime and the position of $\varepsilon^{IC}_{KN}$ with
respect to the characteristic IC frequencies,  we have four cases for each
regime (IC here stands for either EIC and SSC emission).  In the fast cooling
case (valid here for FS-EIC and FS-SSC) we need to solve:

\begin{equation}
{\varepsilon'}_{KN}^{IC}=  \left\{
\begin{array}{lrl}
 (m_e c^2)^2 / \varepsilon'_c & {\rm if} &\gamma_c^2
\varepsilon'_c<{\varepsilon'}_{KN}^{IC}<\gamma_m^2 \varepsilon'_c\\
 m_e c^2 \gamma_m &{\rm if} &\	\gamma_m^2
\varepsilon'_c<{\varepsilon'}_{KN}^{IC}<\gamma_m^2
\varepsilon'_m(=\varepsilon'^{IC}_m)\\
 (m_e c^2)^2/ \varepsilon'_m &{\rm if} &\  \gamma_m^2
\varepsilon'_m<{\varepsilon'}_{KN}^{IC}<\gamma_M^2 \varepsilon'_m\\
 m_e c^2 \gamma_M &{\rm if} &\
\gamma_M^2 \varepsilon'_m <{\varepsilon'}_{KN}^{IC}
\end{array}
\right.
\end{equation}
for ${\varepsilon'}_{KN}^{IC}$. Above the break the spectrum will change to
$F_\varepsilon\propto\varepsilon^{-(p+1-\alpha)}$ or in some cases to
$F_\ve\propto\ve^{-(2-\alpha)}$ in the fast cooling, or to
$F_\ve\propto\ve^{-(p-\alpha)}$ in slow cooling regime.
In the slow cooling regime (used here for RS-SSC and RS-EIC) we can obtain the
break energy by swapping the roles of the cooling ($c$) and characteristic
($m$) quantities in the above equation.

\bibliographystyle{hapj}
\bibliography{pm-all,magnetic,grb,grb-1}
%\bibliography{magnetic,grb}

\end{document}